\documentclass[amssymb,prb,twocolumn,showpacs,floatfix]{revtex4}
\usepackage{amsmath}
\usepackage{bbm}
\usepackage{graphicx}
\usepackage{verbatim}

\begin{document}
\title{\bf
Coherent Spin Rotations in Open Driven Double Quantum Dots.
}
\author{Rafael S\'anchez, Carlos L\'opez-Mon\'is and Gloria Platero}
 \affiliation {Instituto de Ciencia de Materiales, CSIC,
Cantoblanco, Madrid, 28049, Spain.}
\date{\today}
\begin{abstract}
 We analyze coherent spin rotations in a DC biased double
quantum dot driven by crossed DC and AC magnetic fields. In this
configuration, spatial delocalization due to inter-dot tunneling
competes with intra-dot spin rotations induced by the time
dependent magnetic field, giving rise to a complicated time
dependent behavior of the tunneling current. When the Zeeman
splitting has the same value in both dots and spin flip is
negligible, the electrons remain in the triplet subspace
performing coherent spin rotations and current does
not flow. This electronic trapping is removed either by finite
spin relaxation or when the Zeeman splitting is different in each
quantum dot.
In the last case, we will show that applying a resonant
bi-chromatic magnetic field, the electrons become trapped in a
coherent superposition of states and electronic transport is
blocked. Then, manipulating AC magnetic fields allows to drive
electrons to perform  coherent spin rotations which can be
unambiguously detected by direct measurement of the tunneling
current.
\end{abstract}
\maketitle

\section{Introduction}
The accurate tunability of time dependent fields has allowed the access and manipulation of quantum
systems by the resonant illumination of atoms, finding interesting
effects such as the possibility of trapping the atom in a non-absorbing coherent
 superposition ({\it dark state}) which is known as
Coherent Population Trapping
\cite{doubleres,gray,darkst}.
This effect has been applied to non-conducting states in quantum dots (QD) -- also known as {\it artificial
atoms}-- for spinless electrons
\cite{tobiasRenzoni,chu}, having revealed several advantages for
practical issues such as electronic current
switching\cite{tobiasRenzoni} or de-coherence
probing\cite{michaelis}.


Great interest is recently focussed in the coherent control of
electron spin states in the search of candidates for qubits.
Within this scope, optical trapping of localized spins has been
treated in self-assembled quantum dots\cite{economou} and achieved
in diamond deffects\cite{santori}. Electron spin states in QD's
have been proposed as qubits because of their long spin
de-coherence and relaxation times\cite{loss,burkard}. The
controlled
rotation of a single electron spin is one of the challenges for
quantum computation purposes. In combination with the recently
measured controlled exchange gate between two neighboring spins,
driven coherent spin rotations would permit universal quantum
operations.
Recently, experimental and theoretical efforts have been devoted
to describe Electron Spin Resonance (ESR) in single\cite{engel}
and double quantum dots (DQD's)\cite{kopp2,laird}. There, an AC magnetic
field, $B_{\rm AC}$, with a frequency resonant with the Zeeman
splitting $\Delta$ induced by a DC magnetic field, $B_{\rm DC}$,
drives electrons to perform spin coherent rotations which can be
perturbed by electron spin flip induced by scattering processes
such as spin orbit or hyperfine interactions. These are
manifested as a damping of the oscillations. In particular,
hyperfine interaction between electron and nuclei spins induces
flip-flop transitions and an effective Zeeman splitting which adds
to the one induced by $B_{\rm DC}$\cite{naza,fran,kopp2}. ESR mechanism also allows to access spin-orbit physics in the presence of AC electric fields\cite{meier,nowack} or vibrational degrees of freedom in nano-mechanical resonators\cite{lambert}.

In the experiments of Ref. \cite{kopp2}, fast electric field
switching was required in order to reach the Coulomb blockade 
regime and to manipulate the spin electron system. In the present
work we analyze theoretically a simpler configuration, easier to
perform experimentally than the one proposed in \cite{kopp2},
which does not require to bring the double occupied electronic
state in the right dot to the Coulomb blockade configuration and
which consists on conventional tunnel spectroscopy in a DQD under
crossed DC and AC magnetic fields, {\it without additional
electric pulses}. 

The main purpose of this paper is to analyze
the spin dynamics and the tunneling current and to propose for the
first time how to trap electrons in a DQD performing coherent spin
rotations by a resonant AC magnetic field 
which can be unambiguously detected by
conventional tunneling spectroscopy measurements.
We also show how to trap electrons by
means of resonant bi-chromatic magnetic fields in the case where
the Zeeman splitting is different in both QD's (as it usually
happens in the presence of hyperfine interaction).


 We consider a DQD in the {\it spin blockade} 
regime\cite{weinmann},
 i.e., inter-dot tunneling is suppressed due to Pauli exclusion principle\cite{ono} as the electrons in the DQD have
 parallel spins.
This effect may be lifted by the rotation of the electrons spin,
under certain conditions, by the introduction of crossed
 $B_{\rm DC}$ and $B_{\rm AC}$. Then, when $B_{\rm AC}$ is resonant with the Zeeman splitted level, the electrons both rotate their spins within each QD and tunnel, performing
 spatial oscillations between the left and right QD.
The electronic current through such a system performs coherent
oscillations which depend non trivially on both the AC intensity
and the inter-dot coupling.
We will see that, when the effective $B_{\rm DC}$ is
homogeneous through the sample, current is quenched since
 the system is coherently trapped in the triplet subspace ({\it dark subspace}) in spite of the driving field.
However, a finite current may flow as a consequence of spin
relaxation processes.
If $\Delta$ is different within eachQD (it can be due to an
inhomogeneous $B_{\rm DC}$, different g factors or the presence of
hyperfine interaction\cite{naza} with different intensity within each
QD),
$B_{\rm AC}$ is resonant only in one of them and the trapping is lifted.
 Then, off-resonance dynamics of the other electron should in principle affect the total dynamics of the system and it
 should be
included in a theoretical description not restricted to the
Rotating Wave Approximation\cite{hanggi} which is valid just at
resonance. Finally we will show that it is possible to trap the
electrons also in this configuration, where $\Delta$ is different
within each QD, by applying a bichromatic $B_{\rm AC}$, such that each
frequency matches the Zeeman splitting in each QD.

\begin{figure}[t]
\begin{center}
\includegraphics[width=\linewidth,clip]{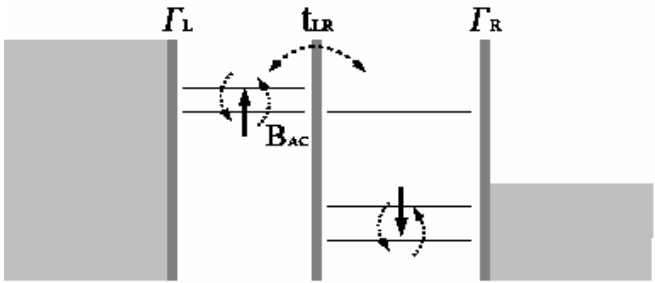}
\end{center}
\caption{\label{esquema}\small Schematic diagram of the DQD in the presence of crossed DC and AC magnetic fields.
}
\end{figure}

\section{Model}\label{model}
Our system consists on two weakly coupled QD's connected to two
fermionic leads, described by the model Hamiltonian: 
\begin{equation}
\hat H(t)=\hat H_0+\hat{H}_{\rm LR}+\hat H_{\rm T}(t)+\hat H_{\rm
leads},
\end{equation}
where $\hat{H}_0=\sum_{i\sigma}\varepsilon_{i}\hat
c_{i\sigma}^\dagger\hat c_{i\sigma}+\sum_{i}U_{i}\hat
n_{i\uparrow}\hat n_{i\downarrow}+V\hat n_{\rm L}\hat n_{\rm R}$
describes the uncoupled DQD, $\hat{H}_{\rm
LR}=-\sum_{\sigma}\left(t_{\rm LR}\hat{c}_{\rm
L\sigma}^\dagger\hat{c}_{\rm R\sigma}+h.c.\right)$ is the
inter-dot coupling and $\hat H_{\rm T}=\sum_{l\epsilon\{\rm
L,R\}k\sigma}(\gamma_{l}\hat{d}_{lk\sigma}^\dagger\hat{c}_{l\sigma}+h.c.)$
gives the tunneling between the DQD and the leads, described by:
$\hat{H}_{\rm leads}=\sum_{lk\sigma}\varepsilon_{lk}\hat
d_{lk\sigma}^\dagger\hat d_{lk\sigma}$, where $\varepsilon_i$ is
the energy of an electron located in dot $i$ and $U_i$ ($V$) is
the intra-dot (inter-dot) Coulomb repulsion. For simplicity, we
disregard the Heisenberg exchange interaction\cite{fran,ono}.
  Finite exchange, would slightly split the inter-dot singlet-triplet energy separation without modifying qualitatively
the results presented here. The chemical potentials of the leads,
$\mu_i$, are such that only two electrons (one in each dot) are
allowed in the system: 
$\varepsilon_i<\mu_i-V<\varepsilon_i+U_i$ and
$\mu_i<\varepsilon_i+2V$. In this configuration, the spin blockade
is manifested when a bias voltage is applied such that the state
with two electrons in the right dot (the one which contributes to
the current) is in resonance with those with one electron in each
dot. The current is then quenched when the electrons in each QD
have the same spin polarization and
Pauli exclusion principle avoids the inter-dot tunneling\cite{ono}.
We now introduce a magnetic field with a
DC component along the $Z$ axis (which breaks the spin
degeneration by a Zeeman splitting $\Delta_i=g_iB_{z,i}$) and a
circularly polarized AC component in the perpendicular plane $XY$
that rotates the $Z$ component of the electron spin when its
frequency satisfies the resonance condition, $\omega=\Delta_i$:
\begin{equation}
\hat{H}_{\rm B}(t)=\sum_i\left[\Delta_iS_z^i+B_{\rm AC}\left(S_x^i\cos\omega
t+S_y^i\sin\omega t\right)\right], 
\end{equation}
where ${\bf S}_i=(1/2)\sum_{\sigma \sigma'} c^\dagger_{i\sigma} {\bf
\sigma}_{\sigma \sigma'} c_{i\sigma'}$ are the spin operators of
each dot (see Fig. \ref{esquema}).

The dynamics of the system is given by the time evolution of the reduced density matrix elements,
 whose equation of motion, within the Born-Markov approximation\cite{blum}, reads:
\begin{eqnarray}
\dot\rho_{ln}(t)&=&-i\langle l|[H_0+H_{\rm LR}+H_{\rm B}(t),\rho]|n\rangle\\
&&+\sum_{k\ne n}\left(\Gamma_{nk}\rho_{kk}-\Gamma_{kn}\rho_{nn}\right)\delta_{ln}-\Lambda_{ln}\rho_{ln}(1-\delta_{ln}).\nonumber
\end{eqnarray}
where the first term in the right hand side accounts for the coherent dynamics within the double quantum dot. $\Gamma_{ln}$ are the transition rates from state $|n\rangle$ to $|l\rangle$ including those induced by the coupling to the leads--being $\Gamma_i=2\pi|\gamma_i|^2$ when they occur through lead $i\epsilon\{{\rm L,R}\}$--and the eventual spin scattering processes (introduced phenomenologically
 by the spin relaxation rate, $T_{1}^{-1}$ ~\cite{prb}). Decoherence appears due to the term $\Lambda_{ln}=\frac{1}{2}\sum_k(\Gamma_{kl}+\Gamma_{kn})+T_2^{-1}$, being ${T_2}=0.1T_{1}$ the intrinsic spin decoherence time. The evolution of the occupation probabilities is
given by the diagonal elements of the density matrix.
In our configuration, the states relevant to the dynamics are:
$|0,\uparrow\rangle$, $|0,\downarrow\rangle$,
$|T_+\rangle=|\uparrow,\uparrow\rangle$,
$|T_-\rangle=|\downarrow,\downarrow\rangle$,
$|\uparrow,\downarrow\rangle$, $|\downarrow,\uparrow\rangle$,
$|S_R\rangle=|0,\uparrow\downarrow\rangle$. This latest state is the only one that contributes to tunneling to the right lead, so the current is given by:  
\begin{equation}
I(t)=2e\Gamma_{\rm R}\rho_{S_{\rm R},S_{\rm R}}(t).
\end{equation} 

Each coherent process is described by a {\it
Rabi-like} frequency. For instance, in the case of two {\it
isolated} spins, one in each QD, which are in resonance with
$B_{\rm AC}$ 
($\Delta_L=\Delta_R$), the oscillation
frequency is: $\Omega_{\rm AC}=2B_{\rm AC}$, see Appendix \ref{appesr}.
On the other hand, the inter-dot tunneling events can be
described by the resonance transitions between the states
$|\uparrow,\downarrow\rangle$, $|\downarrow,\uparrow\rangle$ and
$|S_{\rm R}\rangle$,
whose populations oscillate with a frequency $\Omega_{\rm T}=2\sqrt{2}t_{\rm LR}$, as shown in Appendix \ref{apptun}.

\subsection{$\Delta_{\rm L}=\Delta_{\rm R}$}
\begin{figure}
\begin{center}
\includegraphics[width=\linewidth,clip]{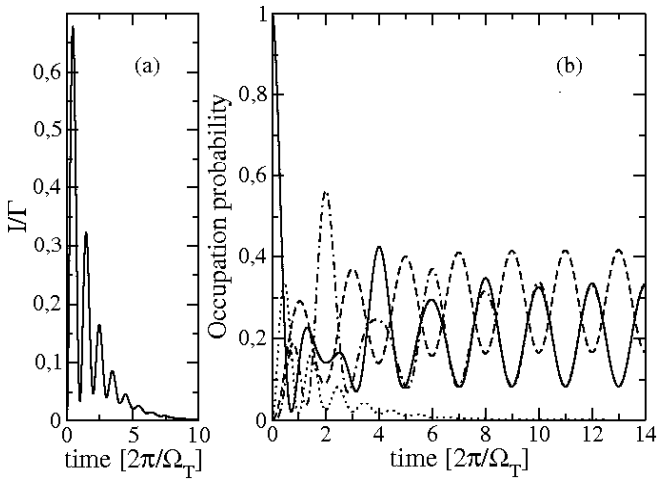}
\end{center}
\caption{\label{ds}\small (a) I(t) for initial state
$|\uparrow,\downarrow\rangle$ in the absence of spin relaxation
for $\Delta_{\rm L}=\Delta_{\rm R}=\Delta$ and $\Omega_{\rm
AC}=\Omega_{\rm T}/2$.  (b) The corresponding occupation
probabilities: $|\uparrow,\downarrow\rangle$ (solid),
$|\downarrow,\uparrow\rangle$ (dash-dotted),
$|0,\uparrow\downarrow\rangle$ (dotted) and
$|\uparrow,\uparrow\rangle$ and $|\downarrow,\downarrow\rangle$
(dashed). Parameters ($e=\hbar=1$): $\Gamma_L=\Gamma_R=\Gamma=10^{-3} {\rm
meV}$, $T_{1(2)}^{-1}=0$, $\Omega_{\rm T}={\rm 11.2 GHz}$ and
holding for the rest of the plots (in meV):
$\varepsilon_{\rm L}=1.5$, $\varepsilon_{\rm R}=0.45$, $\Delta=0.026$ ($B_{\rm DC}\sim1T$), $U_{\rm L}=1$,
$U_{\rm R}=1.45$, $V=0.4$, $\mu_{\rm L}=2$ and $\mu_{\rm R}=1.1$.
}
\end{figure}

We consider initially the case where 
$B_{\rm DC}$ is homogeneous, so that
$\Delta_{\rm R}=\Delta_{\rm L}$ and both spins rotate simultaneously.
Then, the dynamics of the system is properly described in terms of
the dynamics of the total spin of the DQD. $B_{\rm AC}$ acts only
on the states with a finite total magnetic moment: $|T_\pm\rangle$
and
$|T_0\rangle=\frac{1}{\sqrt{2}}(|\uparrow,\downarrow\rangle+|\downarrow,\uparrow\rangle)$,
while the inter-dot tunneling, that does not change the spin, is
only possible between $|S_{\rm R}\rangle$ and
$|S_0\rangle=\frac{1}{\sqrt{2}}(|\uparrow,\downarrow\rangle-|\downarrow,\uparrow\rangle)$.
Therefore, in the absence of spin relaxation, spin rotation
and inter-dot hopping 
are independent processes so any eventual singlet component will
decay by tunneling to the contacts. This produces a finite
current in the transitory regime which drops to zero for longer
times. This process is independent of $B_{\rm AC}$, which is
manifested in the frequency of the current oscillations,
$\Omega_{\rm T}$, cf. Fig. \ref{ds}a. Thus, for large enough times
($t\gg \Gamma_i^{-1}$), transport is cancelled and one electron
will be confined in each QD. The electrons will be coherently
trapped in the inter-dot triplet subspace, $T_\pm$, $T_0$ (dark
subspace) and behave as
an isolated single particle of angular momentum $S=1$ performing
coherent spin rotations with a frequency $\Omega_{AC}$ (Fig. \ref{ds}b).

\begin{figure}[htb]
\begin{center}
\includegraphics[width=\linewidth,clip]{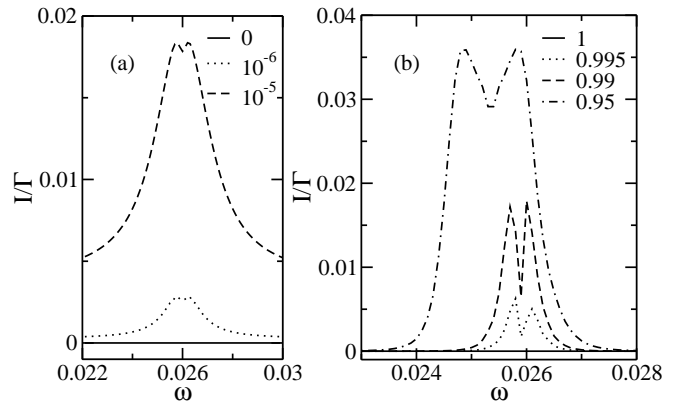}
\end{center}
\caption{\label{DLrelw}\small Effect of (a) finite spin relaxation rates, $T_1^{-1}$ and (b) the Zeeman inhomogenity, $\Delta_{\rm L}/\Delta_{\rm R}$, on the stationary current when tuning the frequency of the magnetic field. In (a), $\Delta_{\rm L}=\Delta_{\rm R}$; in (b), $T_1=0$. (Same parameters as in Fig. \ref{ds} but $\Gamma=10^{-2}{\rm meV}$).
}
\end{figure}

\begin{figure}[t]
\begin{center}
\includegraphics[width=\linewidth,clip]{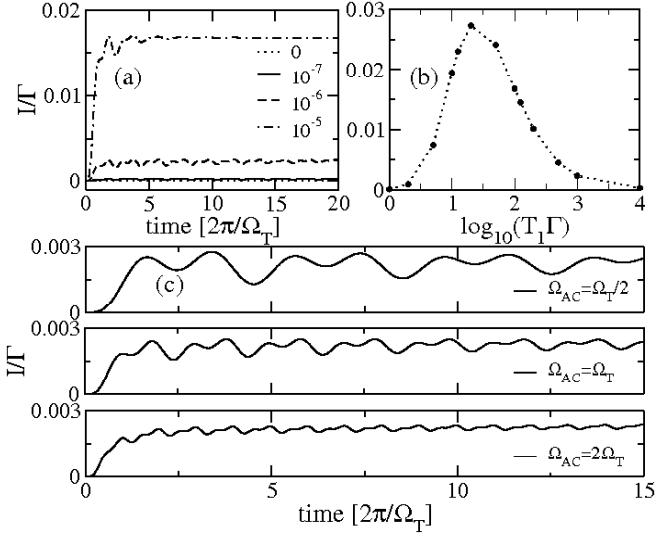}
\end{center}
\caption{\label{irel}\small (a) I(t) for different spin-flip times (in $\mu{\rm s}$), with
 $\Omega_{\rm AC}=\Omega_{\rm T}=11.2  {\rm GHz}$ and $\Delta_{\rm L}=\Delta_{\rm R}=\Delta$. The initial state here is
 $|\uparrow,\uparrow\rangle$, then, for $T_1^{-1}=0$, there is no mixing
  of the triplet and singlet subspaces and therefore, no current flows through the system. Spin relaxation
  processes contribute to populate the singlet, producing a finite current. (b) Stationary
  current as a function of spin relaxation time. For long $T_1$, electrons remain in the dark space. As $T_1$ decreases,
  I begins to flow, being again suppressed for short enough $T_1$, as discussed in the text.
 (c) I(t) for different ratios between the AC field intensity and the inter-dot hopping, i.e.,
between $\Omega_{\rm AC} $ and $\Omega_{\rm T}$, with $T_1\sim 0.1 \mu{\rm s}$. (Same parameters as in Fig. \ref{DLrelw}).
}
\end{figure}

A finite spin relaxation time mixes the dynamics of the singlet
and the triplet subspaces, so that inter-dot tunneling is
allowed and finite current appears, cf. Fig. \ref{DLrelw}a. The shorter the spin relaxation time, the larger is the
singlet-triplet mixing and therefore, the higher is the current,
cf. Fig. \ref{irel}a, up to relaxation times fast enough to
dominate the electron dynamics ($T_1^{-1}\gg\Omega_{\rm AC}$). In this
case, ESR 
is not effective in order to rotate the spins and spin blockade is recovered, cf. Fig. \ref{irel}b.
Since both, spin rotations and spatial delocalization are resonant
processes,
 this singlet-triplet mixing produces complicated dynamics in the current that shows oscillations with a
frequency that depends both on the inter-dot coupling and the AC
field intensity, cf. Fig. \ref{irel}c. 
When $B_{\rm AC}$
increases, the frequency of the current oscillations increases but
not linearly due to the interplay with the hopping. This effect is
small for long spin relaxation times.

\par
\subsection{$\Delta_{\rm L}\ne\Delta_{\rm R}$}
\begin{figure}
\begin{center}
\includegraphics[width=\linewidth,clip]{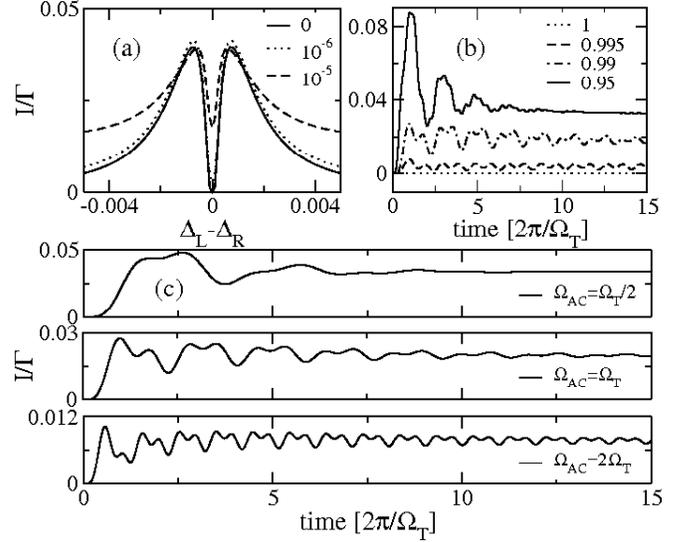}
\end{center}
\caption{\label{izl}\small (a) Dependence of the stationary
current on the DC field inhomogeneity $\Delta_{\rm L}-\Delta_{\rm
R}$  for different relaxation times (in $\mu{\rm s}$). The
quenching of the current for $\Delta_{\rm L}=\Delta_{\rm R}$ is
lifted by spin relaxation. (b) I(t) for different values of
$\Delta_{\rm L}/\Delta_{\rm R}$ when the electron in the right QD
is kept in resonance, in the absence of relaxation. A crossover to
the one electron spin resonance is observed by increasing the
difference between $\Delta_{\rm L}$ and $\Delta_{\rm R}=\Delta$.
(c) Dependence of the current oscillations on $B_{\rm AC}$ for
$\Delta_{\rm L}=0.99\Delta_{\rm R}$ and $T_1^{-1}\sim 0.1\mu{\rm
s}$. Same parameters as in Fig. \ref{irel}.
}
\end{figure}

However, if one introduces an inhomogeneous $ B_{\rm DC}$, so that
only one of the electrons is in resonance with $B_{\rm AC}$ (for
instance, $\omega=\Delta_{\rm R}\ne\Delta_{\rm L}$), the total spin symmetry is broken and then
the electron in each QD behaves differently. In fact, the states
$|\downarrow,\uparrow\rangle$ and $|\uparrow,\downarrow\rangle$
have different occupation probabilities
 and
inter-dot hopping induces
the delocalization of the individual spins. This populates the state
$|S_{\rm R}\rangle$ and a finite current appears showing a double peak whose position shifts following the inhomogenity, cf. Fig. \ref{DLrelw}b. This double peak may be the origin of the under-resolved structure measured in Ref. \cite{kopp2}. By tuning the Zeeman splittings difference, the current presents an
anti-resonance of depth $\sim 0.1{\rm nA}$ near $\Delta_{\rm
L}=\Delta_{\rm R}$, cf. Fig. \ref{izl}a, pretty similar to the coherently trapped atom spectrum in quantum optics\cite{gray}. As expected, taking one
electron slightly out of resonance, the frequency of the current
oscillation is modified in comparison with the double resonance
situation. If one electron is far enough from resonance, the
frequency of the current oscillation becomes roughly half of the
value as it would be the case for the rotation of one electron
spin, cf. Fig. \ref{izl}b. Otherwise, the off-resonant electron
modifies the Rabi frequency for spin rotations in a more
complicated way depending on $B_{\rm AC}$, $t_{\rm LR}$ and  how
much both dynamics are mixed (which is related to $\Delta_{\rm
L}-\Delta_{\rm  R}$), cf. Fig. \ref{izl}c. 
The limiting case when 
$\Delta_{\rm  L}$ and $\Delta_{\rm  R}$ are very different and
only the electron in the right QD is affected effectively by $B_{\rm AC}$ is analyzed in Appendix \ref{appesrtun}.

\section{Bichromatic field}

\begin{figure}
\begin{center}
\includegraphics[width=\linewidth,clip]{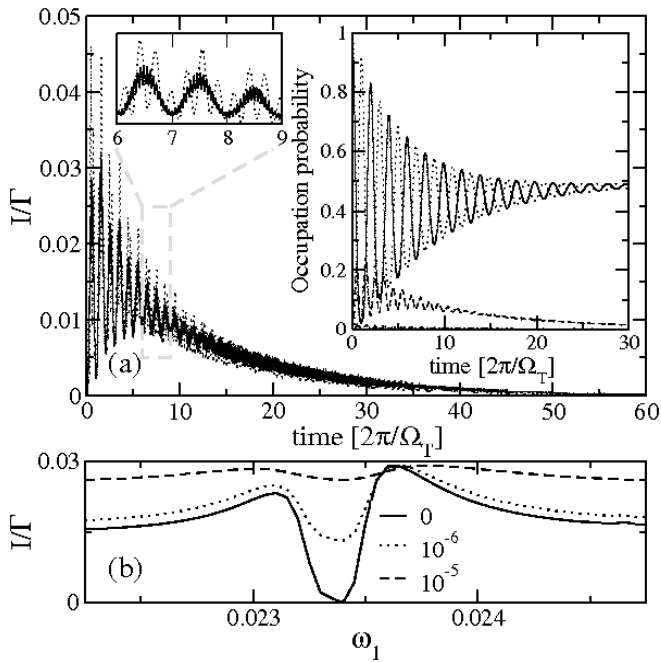}
\end{center}
\caption{\label{2campos}\small (a) Transient current in the
presence of a bichromatic B, when $\omega_{1(2)}=\Delta_{\rm
L(R)}$ for: $\Delta_{\rm L}=\Delta_{\rm R}/2$ (solid) and
$\Delta_{\rm L}=0.9\Delta_{\rm R}$ (dotted) and $T^{-1}_1$=0. Left
inset: detail of the current oscillation. In the case where
$\Delta_{\rm L}=\Delta_{\rm R}/2$, I oscillates with $\Omega_{\rm
T}$ and it presents faster oscillations over-imposed (more
important for $\Delta_{\rm L}=0.9\Delta_{\rm R}$) coming from the
effect of each frequency on its off-resonance electron.
Right Inset: occupation probabilities for $\Delta_{\rm
L}=\Delta_{\rm R}/2$: $|\uparrow,\uparrow\rangle$ (solid),
$|\downarrow,\downarrow\rangle$ (dotted),
$|\uparrow,\downarrow\rangle\sim|\downarrow,\uparrow\rangle$
(dashed) and $|0,\uparrow\downarrow\rangle$ (dash-dotted, remaining very close to zero). The
occupation of $|0,\uparrow\downarrow\rangle$ drops to zero, and
therefore, I drops as well. At long times the electrons fall in a
coherent superposition of $|\uparrow,\uparrow\rangle$ and
$|\downarrow,\downarrow\rangle$. (b) Stationary current
as a function of $\omega_1$ when $\omega_2=\Delta_{\rm R}$, for
different relaxation rates, $T_1^{-1}$. I drops at $\omega_1$
=$\Delta_{\rm L}$. ($\Delta_{\rm R}=\Delta$, $\Gamma=10^{-3}{\rm
meV}$, $\Omega_{\rm T}=1.12{\rm GHz}$).}
\end{figure}

There is a way for trapping the system in a dark state even for
different Zeeman splittings by introducing a bichromatic $B_{\rm
AC}$ with a different frequency that also brings into resonance
the electron in the left QD: 
\begin{equation}\label{hb2}
\hat{H}_{\rm B}^{(2)}(t)=\sum_{{i={\rm L,R}}\atop{j={1,2}}}\left[\Delta_i\hat{S}_z^i+B_{\rm AC}\left(\hat{S}_x^i\cos\omega_j t+\hat{S}_y^i\sin\omega_j t\right)\right],
\end{equation} 
with $\omega_1=\Delta_{\rm L}$ and
$\omega_2=\Delta_{\rm R}$. Then, each electron is resonant with
one of the field frequencies.
In this case, as $\Delta_i$ is different in both QD's, $|T_0\rangle$ mixes
with $|S_0\rangle$ and a finite current flows until the electrons
fall in the superposition:
$|S_2\rangle=\frac{1}{\sqrt{2}}(|\uparrow,\uparrow\rangle-|\downarrow,\downarrow\rangle)$ 
which is not affected by the magnetic field but for
off-resonant oscillations that can be averaged out. In effect, if the non-resonant terms are disregarded, Eq. (\ref{hb2}) is reduced to $\hat{\tilde H}_{\rm
B,0}^{(2)}(t)=\sum_{i}\left[\Delta_i\hat{S}_z^i+B_{\rm AC}\left(\hat{S}_x^i\cos\Delta_it+\hat{S}_y^i\sin\Delta_i t\right)\right]$ and $\hat{\tilde H}_{\rm B,0}^{(2)}|S_2\rangle=0$. Then the
population of the states $|\uparrow,\downarrow\rangle$, 
$|\downarrow,\uparrow\rangle$ and $|S_{\rm R}\rangle$ and, therefore, the current drop to zero, see Fig.~\ref{2campos}a. This transport
quenching also allows to operate the system as a {\it current
switch} by tuning the frequencies of the AC fields (Fig.\ref{2campos}b) and the preparation of the system in a concrete superposition to be manipulated.

The application of a bichromatic magnetic field 
provide a direct measurement of the Zeeman splittings of
the dots
by tuning the frequencies untill the current is brought to a
minimum as in Fig. \ref{2campos}b.
Then, by {\it switching} one of the frequencies off  
and tuning the Zeeman splitting by an additional $B_{\rm DC}$ in one
of the dots, the antiresonance configuration of Fig. \ref{izl}a
could be achieved. In this case, electrons in both QD's perform
coherent spin rotations, as shown in Fig. \ref{ds}b.

\section{Conclusions}
In summary, we present for the first time the complete electron spin dynamics in a 
DQD, in the spin blockade regime, with up to two
extra electrons, where crossed DC and AC magnetic fields and a DC voltage are
applied. In the experimental set up that we propose, different Rabi oscillations 
(due to the ac magnetic field and the interdot tunneling) compete: 
The time dependent magnetic field produces coherent spin
rotations between spin up and down states while resonant inter-dot
hopping allows the spatial delocalization of the electrons. We show how the interplay between 
coherent oscillations coming from inter-dot tunnel and those due to $B_{\rm AC}$ gives rise to
a non trivial electron dynamics which strongly depends on the
ratio between the different Rabi frequencies involved. We show as
well that if $\Delta$ has the same value for the left and the
right QD, electrons remain performing coherent spin rotations in
the $S=1$ subspace and current is quenched. This electron
trapping is removed by spin relaxation or inhomogeneous $B_{\rm
DC}$ and finite current flows. {\it Measuring the current will allow to control coherent spin rotations} and also to
extract information on the spin relaxation time. 
 We propose as well how to block the current by a bichromatic magnetic field in a
 DQD where the effective Zeeman splitting is different within each dot
 (and where current would otherwise flow due to singlet-triplet mixing).
We demonstrate that the bichromatic field induces spin blockade 
in this configuration and that the system evolves to 
a stationary superposition of states, thus serving for spin rectification and state preparation.

Then, our results show that tunneling spectroscopy experiments in
DQD's under tunable mono- and bichromatic magnetic fields allow
to drive electrons to perform coherent spin rotations which
{\it can be unambiguously detected} by measuring the tunneling
current. We  also show how to induce spin blockade in DQDs with different Zeeman splittings by means of a bichromatic magnetic field.

We acknowledge J. I\~narrea and C. Emary for fruitful discussions. This
work has been supported by the MEC (Spain) under grant
MAT2005-06444 and by the EU Marie Curie Network: Project number
504574.

\appendix

\section{Closed system}
In this appendix, we present some simple cases that describe the purely coherent dynamics (i.e. for $\Gamma_{\rm L}=\Gamma_{\rm R}=0$ and $T_1^{-1}=0$) involved in the description presented above. 
\subsection{Two isolated electrons spin resonance}\label{appesr}
We consider first the case where each electron is isolated in one quantum dot.
This system is described by the Hamiltonian $\hat H(t)=\hat H_0+\hat H_{\rm B}(t)$ (as written in Section \ref{model}) and the basis $|1\rangle=|\uparrow,\uparrow\rangle$, $|2\rangle=|\downarrow,\uparrow\rangle$, $|3\rangle=|\uparrow,\downarrow\rangle$ and $|4\rangle=|\downarrow,\downarrow\rangle$. 
We obtain the equations of motion for the reduced density matrix elements from the Liouville equation $\dot\rho(t)=-i[H(t),\rho(t)]$. After a variable transformation: $\rho'_{12,24,34}=e^{-i\omega t}\rho_{12,24,34}$ and $\rho'_{14}=e^{-i2\omega t}\rho_{14}$, they can be written as:
\begin{eqnarray}\label{doseldiag}
\dot\rho_1&=&B_{\rm AC}\Im(\rho'_{21}+\rho'_{31})\nonumber\\
\dot\rho_2&=&B_{\rm AC}\Im(\rho'_{12}+\rho'_{42})\nonumber\\[-2.5mm]
&&\\[-2.5mm]
\dot\rho_3&=&B_{\rm AC}\Im(\rho'_{43}+\rho'_{13})\nonumber\\
\dot\rho_4&=&B_{\rm AC}\Im(\rho'_{34}+\rho'_{24}),\nonumber
\end{eqnarray}
for the diagonal terms, and:
\begin{eqnarray}\label{doseloffdiag}
\dot\rho'_{12}&=&-\frac{i}{2}B_{\rm AC}(\rho_{2}-\rho_{1}+\rho_{32}-\rho'_{14})+i(\Delta_{\rm L}-\omega)\rho'_{12}\nonumber\\
\dot\rho'_{13}&=&-\frac{i}{2}B_{\rm AC}(\rho_{3}-\rho_{1}+\rho_{23}-\rho'_{14})+i(\Delta_{\rm R}-\omega)\rho'_{13}\nonumber\\
\dot\rho'_{14}&=&-\frac{i}{2}B_{\rm AC}(\rho'_{24}+\rho'_{34}-\rho'_{12}-\rho'_{13})+i(\zeta-2\omega)\rho'_{14}\nonumber\\[-2.5mm]
&&\\[-2.5mm]
\dot\rho_{23}&=&-\frac{i}{2}B_{\rm AC}(\rho'_{43}-\rho'_{21}+\rho'_{13}-\rho'_{24})-i\eta\rho_{23}\nonumber\\
\dot\rho'_{24}&=&-\frac{i}{2}B_{\rm AC}(\rho_{4}-\rho_{2}-\rho_{23}+\rho'_{14})+i(\Delta_{\rm R}-\omega)\rho'_{24}\nonumber\\
\dot\rho'_{34}&=&-\frac{i}{2}B_{\rm AC}(\rho_{4}-\rho_{3}-\rho_{32}+\rho'_{14})+i(\Delta_{\rm L}-\omega)\rho'_{34},\nonumber
\end{eqnarray}
for the coherences, where $\zeta=\Delta_{\rm L}+\Delta_{\rm R}$ and $\eta=\Delta_{\rm L}-\Delta_{\rm R}$.

The set of equations (\ref{doseldiag}) and (\ref{doseloffdiag}) can be solved by doing the Laplace transform, ${\cal L}\dot\rho=z\rho-\rho(0)$ and considering the initial condition $\rho_1(0)=1$. If the effect of the magnetic field is the same for both electrons, that is, they suffer the same Zeeman splitting, $\Delta_L=\Delta_R=\Delta$, the probability of finding only one of the electrons flipped, $P_f=\rho_2+\rho_3$ is:
\begin{equation}\label{rhos}
P_f=\frac{2B_{\rm AC}^2}{\Theta^4}\left(\frac{B_{\rm AC}^2}{4}\sin^2\Theta t+\delta^2\sin^2\frac{1}{2}\Theta t\right),
\end{equation} 
where $\Theta^2=B_{\rm AC}^2+\delta^2$ and $\delta=\Delta-\omega$. In the resonant case, $\delta=0$:
\begin{equation}\label{2eres}
P_f=\frac{1}{2}\sin^2B_{\rm AC}t.
\end{equation} 
Therefore, the Rabi frequency for this configuration is:
\begin{equation}\label{rabiB2e}
\Omega_{\rm AC}=2B_{\rm AC},
\end{equation}
twice the one found for the single electron case\cite{cohen}.

On the other hand, if $\Delta_{\rm L}\ne\Delta_{\rm R}$,
the resonance condition holds only for one of them. Then, there is a superposition of different oscillations which results
in a complicated dynamics when $|\eta|\ll\zeta$\cite{genova}.


\subsection{Electron delocalization}\label{apptun}

Let us now consider the closed system in the absence of magnetic field, which can be described by the Hamiltonian $H=H_0+H_{\rm LR}$. The interdot coupling term, $H_{\rm LR}$, induces electron tunneling between both dots, involving the states $|1\rangle=|\uparrow,\downarrow\rangle$, $|2\rangle=|\downarrow,\uparrow\rangle$ and $|3\rangle=|0,\uparrow\downarrow\rangle$. Then, the Liouville equation is given by:
\begin{eqnarray}\label{tunneq}
\dot\rho_1&=&-2t_{\rm LR}\Im\rho_{31}\nonumber\\
\dot\rho_2&=&2t_{\rm LR}\Im\rho_{32}\nonumber\\
\dot\rho_3&=&2t_{\rm LR}\Im(\rho_{31}-\rho_{32})\nonumber\\[-2.5mm]
&&\\[-2.5mm]
\dot\rho_{12}&=&it_{\rm LR}(\rho_{32}+\rho_{13})\nonumber\\
\dot\rho_{13}&=&it_{\rm LR}(\rho_{3}-\rho_{1}+\rho_{12})-i\left(\varepsilon_{\rm L}-\varepsilon_{\rm R}+V-U_{\rm R}\right)\rho_{13}\nonumber\\
\dot\rho_{23}&=&-it_{\rm LR}(\rho_{3}-\rho_{2}+\rho_{21})-i\left(\varepsilon_{\rm L}-\varepsilon_{\rm R}+V-U_{\rm R}\right)\rho_{23}\nonumber
\end{eqnarray}

By solving the set of equations (\ref{tunneq}) under the condition $\varepsilon_{\rm L}-\varepsilon_{\rm R}=U_{\rm R}-V$, where interdot tunneling is resonant, we obtain the occupation of the state $|0,\uparrow\downarrow\rangle$, which is given by:
\begin{equation}
\rho_3=\frac{1}{2}\sin^2\sqrt{2}t_{\rm LR}t.
\end{equation}
Thus, the Rabi frequency is modified respect to the single electron case ($\Omega_{1e}=2t_{\rm LR}$\cite{cohen}):
\begin{equation}\label{rabitun2e}
\Omega_{\rm T}=2\sqrt 2t_{\rm LR}.
\end{equation}

\subsection{Mixing of spatial delocalization and spin rotation}\label{appesrtun}
As discussed in the text for the open system, if $\Delta_{\rm L}\ne\Delta_{\rm R}$, the current shows a coherent oscillation that depends in both the intensity of the AC magnetic field and the interdot hopping. 

Here we consider a simple case that presents both coherent processes --spin rotation and interdot delocalization-- by considering very different Zeeman splittings in each QD. Then, only one of the electrons is in resonance with the AC field: $\Delta_{\rm L}=\omega$ and only three states contribute to the dynamics: $|1\rangle=|\uparrow,\uparrow\rangle$, $|2\rangle=|\downarrow,\uparrow\rangle$ and $|3\rangle=|0,\uparrow\downarrow\rangle$, resulting in the set of equations:
\begin{eqnarray}
\dot\rho_1&=&B_{\rm AC}\Im\rho_{21}\nonumber\\
\dot\rho_2&=&B_{\rm AC}\Im\rho_{12}+2t_{\rm LR}\Im\rho_{32}\nonumber\\
\dot\rho_3&=&-2t_{\rm LR}\Im\rho_{32}\nonumber\\[-2.5mm]
&&\\[-2.5mm]
\dot\rho_{12}&=&-\frac{i}{2}B_{\rm AC}(\rho_{2}-\rho_{1})+it_{\rm LR}\rho_{13}+i(\Delta_{\rm L}-\omega)\rho_{12}\nonumber\\
\dot\rho_{13}&=&-\frac{i}{2}B_{\rm AC}\rho_{23}+it_{LR}\rho_{12}-i\omega_+\rho_{13}\nonumber\\
\dot\rho_{23}&=&-\frac{i}{2}B_{\rm AC}\rho_{13}-it_{\rm LR}(\rho_3-\rho_{2})-i\omega_-\rho_{23},\nonumber
\end{eqnarray}
where $\omega_\pm=\varepsilon_{\rm L}-\varepsilon_{\rm R}-\frac{\Delta_{\rm R}\pm\Delta_{\rm L}}{2}+V-U_{\rm R}$.
If the gate voltages are tuned in a way that $\omega_-=0$, so the left electron can tunnel to the doubly occupied singlet state in the right dot (having both electrons opposite spin polarization), one finds that the frequency of the oscillations depends on both the tunneling coupling and the field intensity: $\Omega\propto\sqrt{B_{\rm AC}^2+4t_{\rm LR}^2}$.

\end{document}